\documentclass[twocolumn,showpacs,preprintnumbers,amsmath,amssymb,times]{revtex4}
\usepackage{amsfonts}
\usepackage{graphicx}
\usepackage{dcolumn}
\usepackage{bm}
\usepackage{amsmath}

\begin{document}
\title{Epidemic spreading and immunization with identical infectivity}
\author{Rui Yang$^1$}
\author{Jie Ren$^2$}
\author{Wen-Jie Bai$^1$}
\author{Tao Zhou$^{1,2}$}
\email{zhutou@ustc.edu}
\author{Ming-Feng Zhang$^1$}
\author{Bing-Hong Wang$^1$}

\affiliation{$^1$Department of Modern Physics and Nonlinear
Science Center, University of Science and Technology of China,
Hefei 230026, P. R. China \\ $^2$Department of Physics, University
of Fribourg, CH-1700 Fribourg, Switzerland}

\date{\today}

\begin{abstract}
In this paper, a susceptible-infected-susceptible (SIS) model with
identical infectivity, where each node is assigned with the same
capability of active contacts, $A$, at each time step, is
presented. We found that on scale-free networks, the density of
the infected nodes shows the existence of threshold, whose value
equals $1/A$, both demonstrated by analysis and numerical
simulation. The infected population grows in an exponential form
and follows hierarchical dynamics, indicating that once the highly
connected hubs are reached, the infection pervades almost the
whole network in a progressive cascade. In addition, the effects
of random, proportional, and targeted immunization for this model
are investigated. Based on the current model and for heterogenous
networks, the targeted strategy performs best, while the random
strategy is much more efficient than in the standard SIS model.
The present results could be of practical importance in the setup
of dynamic control strategies.
\end{abstract}


\maketitle
\section{Introduction}
Epidemic, one of the most important issues related to our real
lives, such as computer virus on Internet and venereal disease on
sexual contact networks, attracts a lot of attention. Among all the
models on the process of the epidemic, susceptible-infected (SI)
model \cite{SI1,SI2}, susceptible-infected-susceptible (SIS) model
\cite{SIS1,SIS2}, and susceptible-infected-removed (SIR) model
\cite{SIR1,SIR2,SIR3}, are considered as the theoretical templates
since they can, at least, capture some key features of real
epidemics. After some classical conclusions have been achieved on
regular and random networks, recent studies on small-world (SW)
networks \cite{Watts1998} and scale-free (SF) networks \cite{BA}
introduce fresh air into this long standing area (see the reviews
\cite{Reviews} and the references therein). The most striking result
is that in the SIS and SIR model, the critical threshold vanishes in
the limit of infinite-size SF networks. It is also a possible
explanation why some diseases are able to survive for a long time
with very low spreading rate.

In this paper, we focus on the SIS model. Although it has achieved
a big success, the standard SIS style might contain some
unexpected assumption while being introduced to the SF networks
directly, that is, each node's potential infection-activity
(infectivity), measured by its possibly maximal contribution to
the propagation process within one time step, is strictly equal to
its degree. As a result, in the SF networks the nodes with large
degree, named \emph{hubs}, will take the greater possession of the
infectivity, so-called \emph{super-spreader}. This assumption may
fail to mimic some cases in the real world where the relation
between degree and infectivity is not simply equal
\cite{relation}. The first example is that, in most of the
existing peer-to-peer distributed systems, although their
long-term communicating connectivity shows the scale-free
characteristic \cite{p1}, all peers have identical capabilities
and responsibilities to communicate at a short term, such as the
Gnutella networks \cite{p2}. Second, in sexual contact networks ,
even the hub node has many acquaintances; he/she has limited
capability to contact with others during limited periods
\cite{SCN}. Third, the referral of a product to potential
consumers costs money and time in network marketing processes
(e.g. a salesman has to make phone calls to persuade his social
surrounding to buy the product). Therefore, the salesman will not
make referrals to all his acquaintances \cite{market}. The last
one, in some email service systems, such as the Gmail system
schemed out by Google \cite{google}, the clients are assigned by
limited capability to invite others to become Gmail-user after
being invited by an E-mail from another Gmail-user. Similar
phenomena are common in our daily lives, thus need a further
investigation.

\section{The model}
In the epidemic contact network, node presents individual and link
denotes the potential contacts along which infections can spread.
Each individual can be in two discrete states, whether susceptible
(S) or infected (I). At each time step, the susceptible node which
is connected to the infected one will be infected with rate
$\beta$. Meanwhile, infected nodes will be cured to be again
susceptible with rate $\delta$, defining the effective spreading
rate as $\lambda=\beta/\delta$. Without losing of generality, we
set $\delta=1$. Individuals run stochastically through the cycle
susceptible-infected-susceptible, which is also the origin of the
name, SIS. Denote $S(t)$ and $I(t)$ the density of the susceptible
and infected population at the time step $t$, respectively. Then
\begin{equation}
I(t)+S(t) = 1.
\end{equation}

In the standard SIS model, each individual will contact all its
neighbors once at each time step, thus the infectivity of each node
is equal to its degree. In the present model, we assume that every
individual has the same infectivity $A$. That is to say, at each
time step, each infected individual will generate $A$ contacts where
$A$ is a constant. Multiple contacts to one neighbor are allowed,
and the contacts to the infected ones, although without any effect
on the epidemic dynamics, are also counted. In this paper, with half
nodes infected initially, we run the spreading process for
sufficiently long time, and calculate the fraction of infected nodes
averaging over the last 1000 steps as the density of infected nodes
in the steady stage (denoted by $\rho$). All of our simulation
results are obtained from averaging over $300$ different network
realizations, and for each $100$ independent runs with different
initial configurations.

\begin{figure}
\scalebox{0.7}[0.7]{\includegraphics{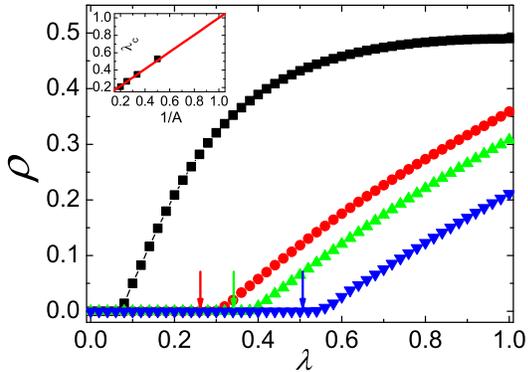}}
\caption{\label{fig:epsart} (Color online) Average value of $\rho$
as a function of the effective spreading rate $\lambda$ on a BA
network with average degree $\langle k\rangle=8$ and network size
$N=2000$. The black points represent the case of standard SIS
model, and the red, green and blue points correspond to the
present model with $A=4$, 3 and 2, respectively. The arrows point
at the critical points gained from the simulation. The insert
shows the threshold $\lambda_c$ scaling with $1/A$, with solid
line representing the analytical results.}
\end{figure}

\section{Threshold behavior}
Let $I_k(t)$ denote the fraction of vertices of degree $k$ that
are infected at time $t$. Then using the mean-field approximation,
the rate equation for the partial densities $I_k(t)$ in a network
characterized by a degree distribution $P(k)$ can be written as:
\begin{equation}
\partial_t I_k(t)=-I_k(t)+\lambda
k[1-I_k(t)]\sum_{k'}\frac{P(k'|k) I_{k'}(t)A}{k'},
\end{equation}
where $P(k'|k)$ denotes the conditional probability that a vertex
of degree $k$ is connected to a vertex of degree $k'$. Considered
the uncorrelated networks, where $P(k'|k)=k'P(k')/\langle
k\rangle$, the rate equation takes the form:
\begin{equation}
\partial_t I_k(t)=-I_k(t)+\lambda\frac{k}{\langle
k\rangle}[1-I_k(t)]I(t)A.
\end{equation}
Using $\rho_k$ to denote the value of $I_k(t)$ in the steady stage
with sufficiently large $t$, then
\begin{equation}
\partial_t \rho_k=0,
\end{equation}
which yields the nonzero solutions
\begin{equation}
\rho_k=\frac{\lambda k\rho A/\langle k\rangle}{1+\lambda k\rho
A/\langle k\rangle},
\end{equation}
where $\rho=\sum_{k}P(k)\rho_{k}$ is the infected density at the
network level in the steady stage. Then, one obtains
\begin{equation}
\rho=\frac{\lambda \rho A}{\langle
k\rangle}\sum_{k}\frac{kP(k)}{1+A\lambda k\rho/\langle k\rangle}.
\end{equation}
To the end, for the critical point where $\rho\sim0$, we get
\begin{equation}
\lambda A=\frac{\langle k\rangle }{\sum_{k}kP(k)}=1 .
\end{equation}
This equation defines the epidemic threshold
\begin{equation}
\lambda_c=\frac{1}{A},
\end{equation}
below which the epidemic prevalence is null, and above which it
attains a finite value. The previous works about epidemic
spreading in SF networks present us with a completely new scenario
that a highly heterogeneous structure will lead to the absence of
any epidemic threshold \cite{Reviews}, while now, in the present
model, it is $1/A$ instead. As shown in Fig. 1, the analytical
result agrees very well with the simulations. Furthermore, it is
also clear that the larger infectivity $A$ will lead to the higher
prevalence $\rho$.

\begin{figure}
\scalebox{0.65}[0.65]{\includegraphics{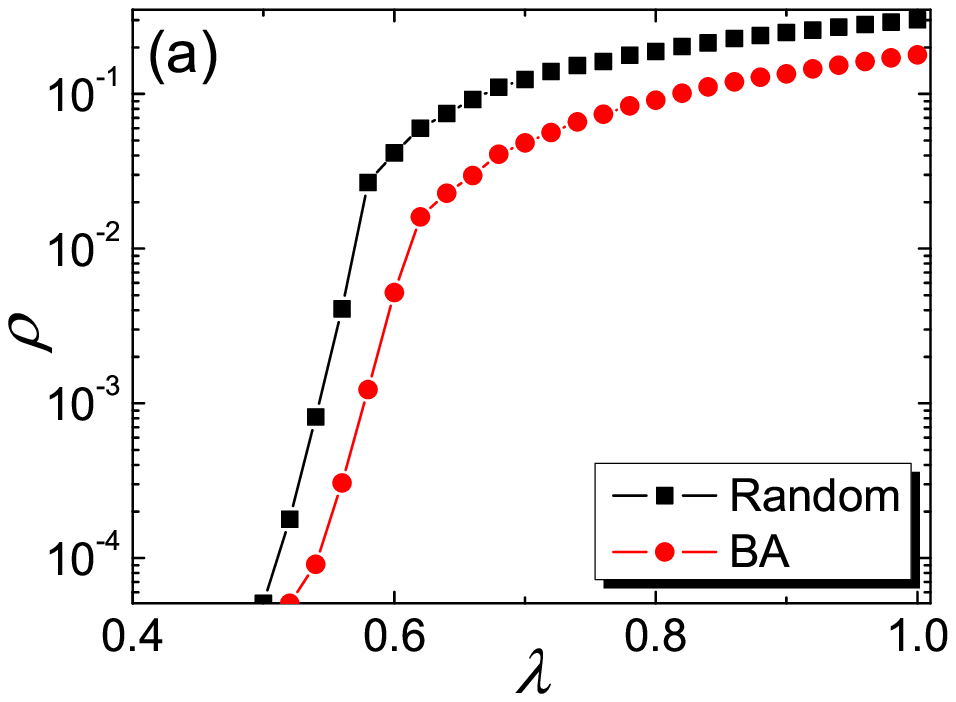}}
\scalebox{0.65}[0.65]{\includegraphics{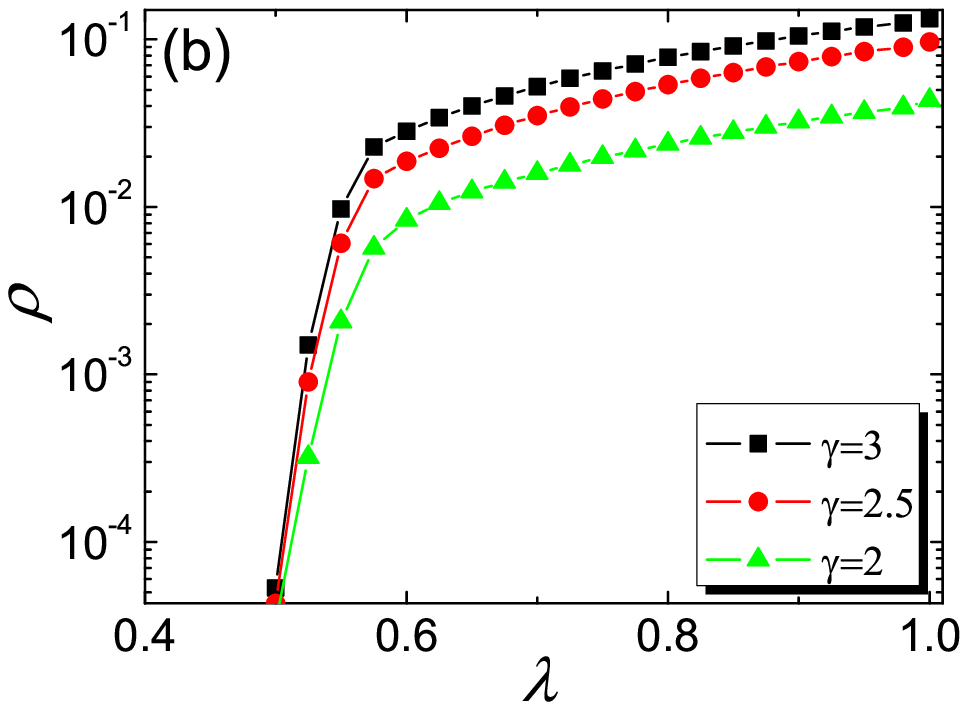}}
\scalebox{0.65}[0.65]{\includegraphics{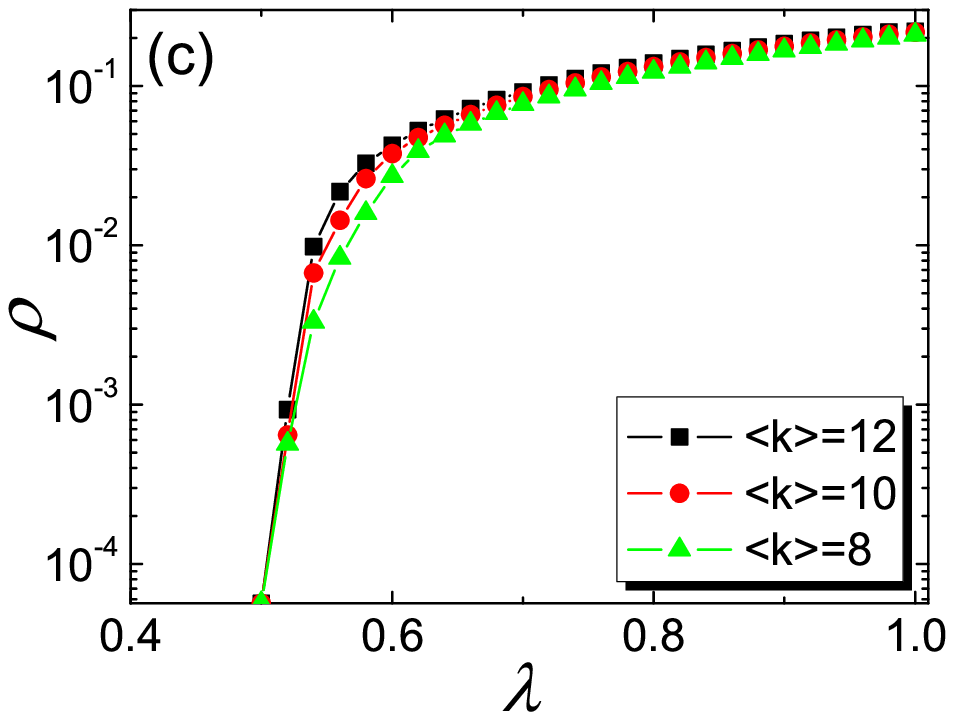}}
\caption{\label{fig:epsart}(Color online) Average value of $\rho$
as a function of the effective spreading rate $\lambda$ on (a) the
random and BA networks; (b) the SF configuration networks for
different values of $\gamma$; (c) the BA networks with different
average degree. In (a) and (b), the average degree is with
$\langle k\rangle=6$, and for all the simulations, $N=2000$ and
$A=2$ are fixed.}
\end{figure}

From the analytical result of the threshold value,
$\lambda_c=1/A$, we can also acquire that the critical behavior is
independent of the topology of networks which are valid for the
mean-field approximation \cite{reason}. To demonstrate this
proposition, we implement the present model on various networks;
These include the random networks, the scale-free configuration
model \cite{EPJB1} with different power-law exponent $\gamma$, and
the BA networks with different average degree. As shown in Fig. 2,
under a given $A$, the critical value are the same, which strongly
support the valid of Eq. (8). Furthermore, there is no distinct
finite-size effect as shown in Fig. 3. In the original SIS model,
the node's infectivity relies strictly on its degree $k$ and the
threshold is $\lambda_c \sim \langle k\rangle/\langle k^2\rangle$.
Since the variance of degrees gets divergent with the increase of
$N$, the epidemic propagation on scale-free networks has an
obvious size effect \cite{finite-size}. However, in the current
model, each infected node is just able to contact the same number
of neighbors, $A$, rather than its degree. Thus the threshold
value and the infected density beyond the threshold are both
independent of the size $N$.

\begin{figure}
\scalebox{0.7}[0.7]{\includegraphics{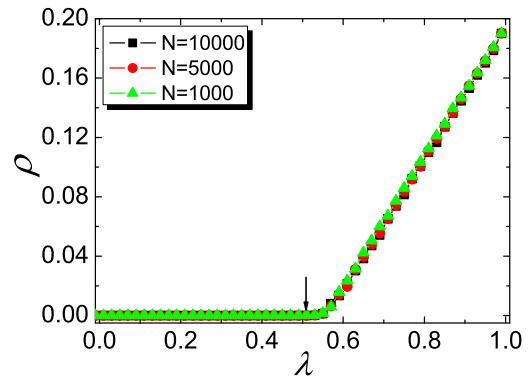}}
\caption{\label{fig:epsart}(Color online) Average value of $\rho$
as a function of the effective spreading rate $\lambda$ on the
different sizes of BA networks with $\langle k\rangle=6$ and
$A=2$.}
\end{figure}

\section{Time behavior}
For further understanding of the epidemic dynamics of the proposed
model, we study the time behavior of the epidemic propagation.
First of all, manipulating the operator $\sum_{k}{P(k)}$ on both
sides of Eq. (3), and neglecting the terms of order
$\mathbb{O}(I^2)$, we obtain
\begin{equation}
\partial_t I(t)=-I(t)+\lambda AI(t).
\end{equation}
Thus the evolution of $I(t)$ follows an exponential growing as
\begin{equation}
I(t)\sim e^{ct},
\end{equation}
where $c\varpropto (\lambda A-1)$.

\begin{figure}
\scalebox{0.7}[0.7]{\includegraphics{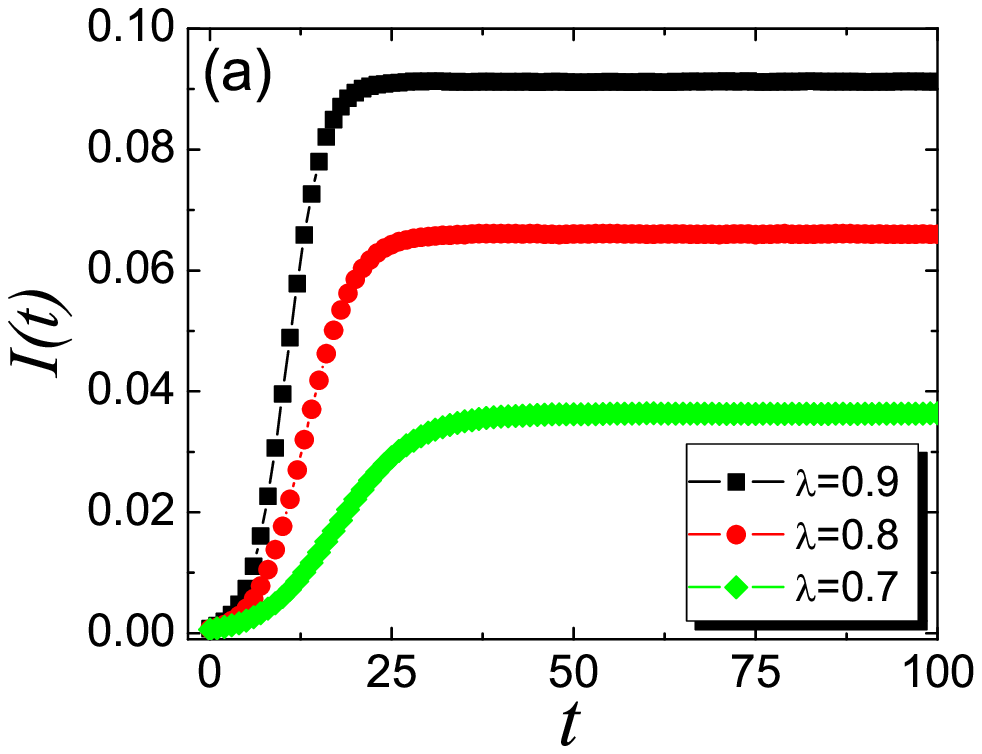}}
\scalebox{0.7}[0.7]{\includegraphics{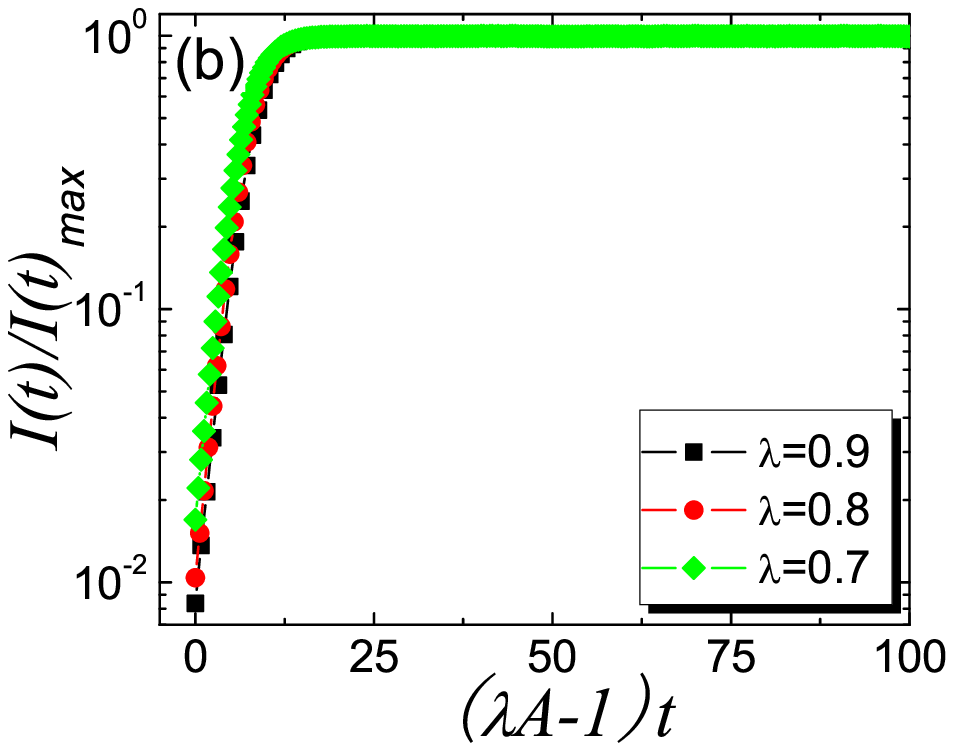}}
\caption{\label{fig:epsart}(Color online) Average value of $I(t)$
in normal plots as time $t$ (a) and $I(t)/{I(t)_{max}}$ in
single-log plots as rescaled time $(\lambda A-1)t$ (b) for
different spreading rate $\lambda$. The numerical simulations are
implemented based on BA networks of size $N=2000$, $\langle
k\rangle=6$, and $A=2$.}
\end{figure}

In Fig. 4, we report the simulation results of the present model
for different spreading rates ranging from 0.7 to 0.9. The
rescaled curves $I(t)/{I(t)_{max}}$ (Fig. 4(b)) can be well fitted
by a straight line in single-log plot for small $t$ and the curves
corresponding to different $\lambda$ will collapse to one curve
with rescaling time $(\lambda A-1)t$, which strongly supports the
analytical result Eq. (10).

Furthermore, a more precise characterization of the epidemic
diffusion through the network can be achieved by studying some
convenient quantities in numerical experiments. First, we measure
the average degree of newly infected nodes at time $t$ as
\begin{eqnarray}
\langle k_{inf}{(t)} \rangle=\frac{\sum{_k}{k I_k(t)}}{I(t)}.
\end{eqnarray}
Then, we present the inverse participation ratio $Y_2(t)$ to
indicate the detailed information on the infection propagation,
which is defined as \cite{21}:
\begin{equation}
Y_2(t)=\sum_{k}{w_k^2(t)},
\end{equation}
where the weight of recovered individuals in each $k$-degree class
(here $k$-degree class means the set of all the nodes with degree
$k$) is defined by $w_k(t)=I_{k}(t)/I(t)$. From this definition,
one can acquire that if $Y_2$ is small, the infected are
homogeneously distributed among all degree classes; on the
contrary, if $Y_2$ is relatively larger then, the infection is
localized on some specific degree classes.

\begin{figure}
\scalebox{0.7}[0.7]{\includegraphics{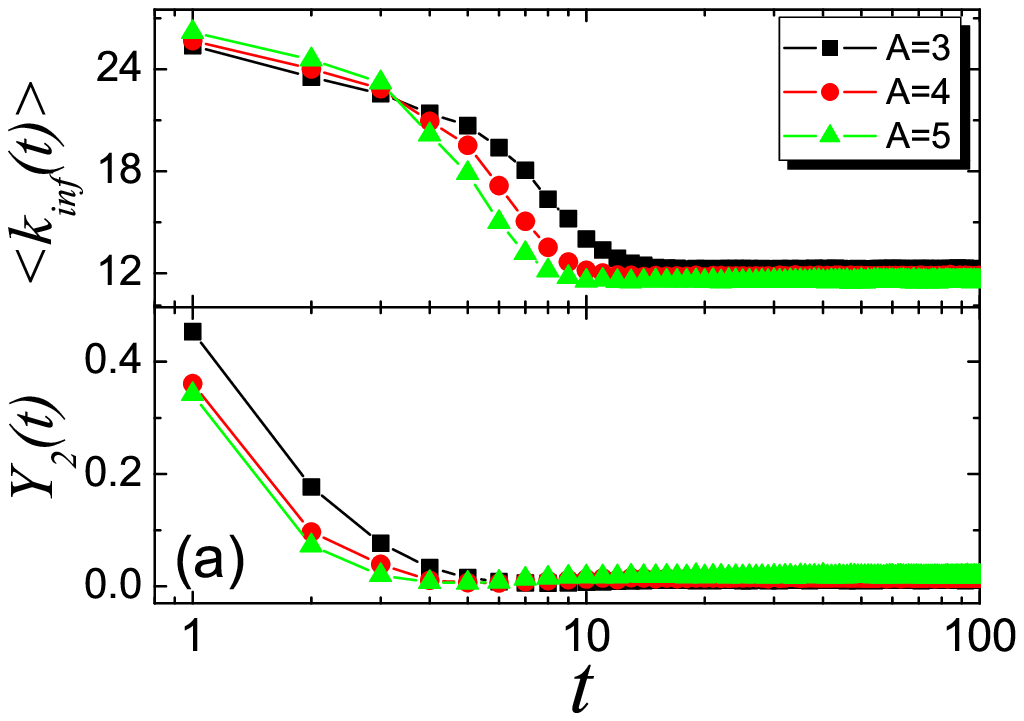}}
\scalebox{0.7}[0.7]{\includegraphics{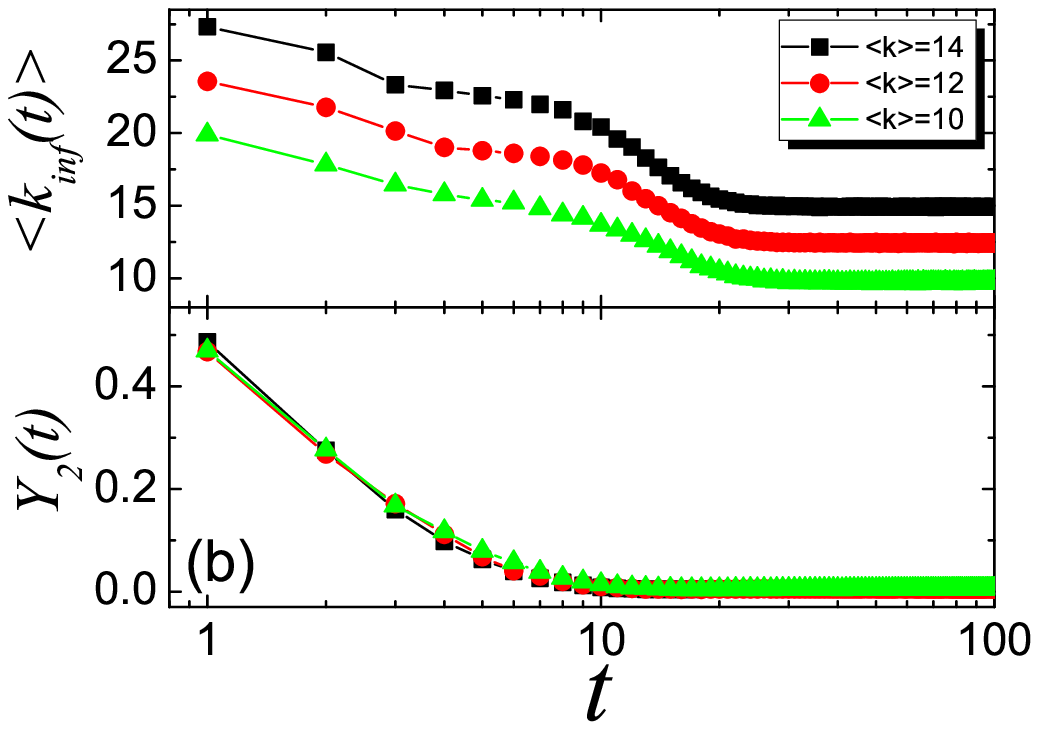}}
\caption{\label{fig:epsart}(Color online) Time behavior of the
average degree of the newly infected nodes (top) and inverse
participation ratio $Y_2$ (bottom) in BA networks of size $N=2000$,
$\lambda=0.8$ for different values of $A$ (with $\langle
k\rangle=12$ fixed) (a) and $\langle k\rangle$ (with $A=2$ fixed)
(b).}
\end{figure}

In Fig. 5, we exhibit the time behaviors of these quantities for
BA networks and find a hierarchical dynamics, that is, all those
curves show an initial plateau, which denotes that the infection
takes control of the large degree nodes firstly. Once the highly
connected hubs are reached, the infection pervades almost the
whole network via a hierarchical cascade across smaller degree
classes. Thus, $\langle k_{inf}{(t)}\rangle$ decreases to the next
plateau, which approximates the average degree $\langle k
\rangle$.

\section{Immunization}
Immunity, relating to the people's strategies to struggle with the
disease epidemics, shows great importance in practice
\cite{Reviews}. Since the current model, which can mimic some real
cases more accurately, shows different characters with the
standard SIS model, it requires some in-depth and detailed
investigation about the immunity on this model. As we know,
immunized nodes cannot become infected and, therefore, will not
transmit the infection to their neighbors. The simplest
immunization strategy is to select immunization population
completely randomly, so-called \emph{random immunization}
\cite{random}. However, this strategy is inefficient for
heterogenous networks. Similar to the preferential attachment
mechanism introduced by BA model \cite{BA}, Dezs\"{o} and
Barab\'{a}si proposed the \emph{proportional immunization}
strategy \cite{proportional}, in which the immunizing probability
of each node is proportional to its degree. This preferential
selection strategy can remarkable enhance the immunization
efficiency in scale-free networks. The extreme strategy for
immunization in heterogenous networks is the so-called
\emph{targeted immunization} \cite{target}, where the most highly
connected nodes are chosen to be immunized. Compared with the
random immunization and proportional immunization, the targeted
immunization is demonstrated as the most efficient one for various
networks \cite{target-net}, and several different but relative
dynamics \cite{target-model}.

\begin{figure}
\scalebox{0.7}[0.7]{\includegraphics{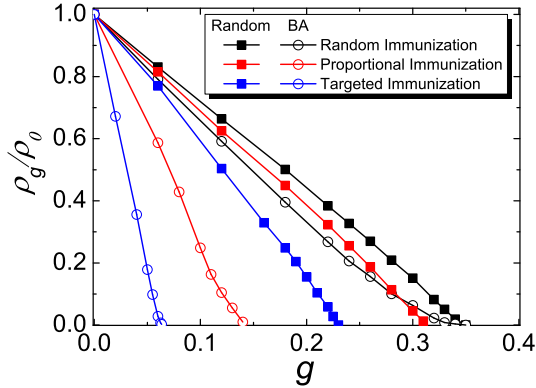}}
\caption{\label{fig:epsart}(Color online) Reduced prevalence
$\rho_g/\rho_0$ from numerical simulations of the present model in
the random (square point) and BA (circle point) network with
random (black line), proportional (red line) and targeted
immunizations (blue line). In the simulations, the parameter
$\lambda=0.8$, $A=2$, $\langle k\rangle=6$ and $N=2000$ are
fixed.}
\end{figure}

In Fig. 6, we report the simulation results about the three
mentioned immunization strategies on the current model. The
$x$-axis, $g$, denotes the fraction of immunized population, and
the $y$-axis, $\rho_g/\rho_0$, represents the performance, where
$\rho_0$ is the prevalence of infected nodes without immunization
and $\rho_g$ the one after immunization. From the simulation
results, one can find that the epidemic thresholds under random,
proportional and targeted immunizations of random networks are
$g_c\simeq0.35$, $0.32$ and $0.23$, respectively. And those of BA
networks are $g_c\simeq0.35$, $0.14$ and $0.07$. It is clear from
the simulation results, even in the current model where the
infectivities of large-degree nodes are greatly suppressed, the
targeted immunization performs best. Combine with the hierarchical
behavior observed in Sec. IV, it strongly indicates that the
heterogeneities of degree and infectivity could both contribute to
the violent spreading of disease. Hence even for the current model
with identical infectivity, the hub nodes play much more important
roles in determining the dynamical property.

Note that, in this model, for heterogenous networks, the random
immunization is more efficient than the standard case, and the
threshold $g_c$ is the same for BA and random networks. Actually,
the random immunization is implemented by randomly selecting and
immunizing $gN$ nodes on a network of fixed size $N$. At the
mean-field level, the presence of uniform immunity will effectively
reduce the spreading rate $\lambda$ by a factor $(1-g)$. According
to Eq. (8), the immunization threshold is given by
\begin{equation}
g_c=1-\frac{1}{A\lambda}.
\end{equation}
As shown in Fig. 6, the simulated result ($g_c \simeq 0.35$)
agrees with the analytical result ($g_c=0.375$) well. To compare,
the random immunization threshold of standard SIS model is given
by $g_c=1-{\langle k\rangle}/{\lambda\langle k^2\rangle}$
\cite{target}. Namely, to control the spreading, one have to
immunize all the population as $g_c(N)\rightarrow 1$ in the
thermodynamic limit $N\rightarrow \infty$.

\begin{figure}
\scalebox{0.7}[0.7]{\includegraphics{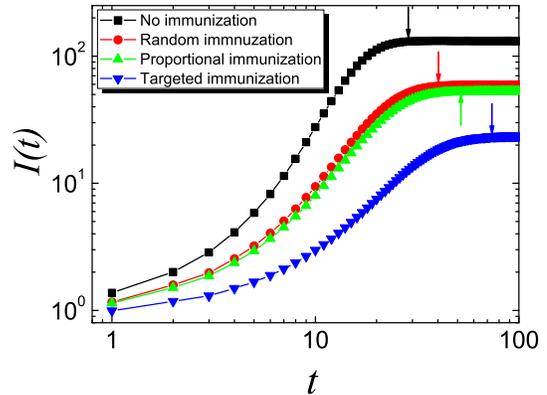}}
\caption{\label{fig:epsart}(Color online) Average value of $I(t)$
as time $t$ for no immunization (black), random immunization
(red), proportional immunization (green), and targeted
immunization (blue) at $\lambda=0.8$ and $g=0.01$. The numerical
simulations are implemented based on BA networks of size $N=2000$,
$\langle k\rangle=6$, and $A=2$. The arrows indicate the time that
the whole spreading process comes to the steady stage.}
\end{figure}

For further understanding the effects of those different
immunization strategies, we study the time behaviors as shown in
Fig. 7. In accordance with the above results, the spreading
velocity under target immunization is the lowest. Note that,
different from the standard SIS model, the random immunization can
obviously slow down the spreading in the early stage even with a
tiny population $g \sim 1\%$.

\section{Conclusion}
In this paper, we investigated the behaviors of SIS epidemics with
the identical infectivity $A$. By comparing the dynamical
behaviors of the present model of different values of $A$ with the
standard one on BA networks, we found the existence of epidemic
spreading threshold. The analytical result of the threshold $1/A$
is provided, which agrees with numerical simulation very well. The
critical value is independent of the topology of underlying
networks, just depends on the dynamical parameter $A$ and the
whole spreading process does not have the distinct finite-size
effect. For SF networks, the infected population grows in an
exponential form in the early stage, and then follows a
hierarchical dynamics. In addition, the time scale is also
independent of the underlying topology.

The last but not the least, the numerical results of random,
proportional, and targeted immunization are presented. We found
that the targeted immunization performs best, while the random
immunization is much more efficient in heterogenous networks than
the standard case.

\begin{acknowledgments}
BHWang acknowledges the support of 973 Project under Grant No.
2006CB705500, the Special Research Founds for Theoretical Physics
Frontier Problems under Grant No. A0524701, the Specialized
Program under the Presidential Funds of the Chinese Academy of
Science, and the National Natural Science Foundation of China
under Grant No. 10472116. TZhou acknowledges the support of the
National Natural Science Foundation of China under Grant Nos.
70471033 and 10635040.
\end{acknowledgments}


\begin{thebibliography}{ref1}

\bibitem{SI1} M. Barth\'elemy, A. Barrat, R. Pastor-Satorras, and A. Vespihnani, Phys. Rev. Lett. \textbf{92}, 178701
(2004); M. Barth\'elemy, A. Barrat, R. Pastor-Satorras, and A.
Vespihnani, J. Theor. Biol. {\bf 235}, 275 (2005).

\bibitem{SI2} T. Zhou, G. Yan, and B. -H. Wang, Phys. Rev. E {\bf 71}, 046141
(2005); G. Yan, T. Zhou, J. Wang, Z. -Q. Fu, and B. -H. Wang,
Chin. Phys. Lett. {\bf 22}, 510 (2005).

\bibitem{SIS1} R. Pastor-Satorras and A. Vespignani, Phys. Rev.
Lett. \textbf{86}, 3200 (2001); R. Pastor-Satorras and A.
Vespignani, Phys. Rev. E \textbf{63}, 066117 (2001).

\bibitem{SIS2} M. Bogu\~{n}\'{a}, and R. Pastor-Satorras, Phys.
Rev. E {\bf 66}, 047104 (2002); M. Bogu\~{n}\'{a}, R.
Pastor-Satorras, and A. Vespignani, Phys. Rev. Lett. {\bf 90},
028701 (2003).

\bibitem{SIR1} R. M. May and A. L. Lloyd, Phys. Rev. E \textbf{64}, 066112
(2001); A. L. Lloyd, and R. M. May, Science {\bf 292}, 1316
(2001).

\bibitem{SIR2} Y. Moreno, R. Pastor-Satorras
and A. Vespignani, Eur. Phys. J. B \textbf{26}, 521 (2002); Y.
Moreno, J. B. Gomez, and A. F. Pacheco, Phys. Rev. E {\bf 68},
035103 (2003).

\bibitem{SIR3} G. Yan, Z. -Q. Fu, J. Ren, W. -X. Wang, Phys. Rev. E {\bf 75},
016108 (2007).

\bibitem{Watts1998} D. J. Watts and S. H. Strogats, Nature
(london) \textbf{393}, 440 (1998).

\bibitem{BA} A.-L. Barab\'asi and R. Albert, Science
\textbf{286}, 509 (1999).

\bibitem{Reviews} R. Pastor-Satorras, and A. Vespignani, \emph{Epidemics and immunization in scale-free networks}. In: S. Bornholdt, and H. G. Schuster (eds.) \emph{Handbook of Graph and Networks}, Wiley-VCH, Berlin, 2003; T. Zhou, Z. -Q. Fu, and B. -H. Wang, Prog. Nat. Sci. {\bf 16}, 452
(2006); S. Boccaletti, V. Latora, Y. Moreno, M. Chavez, and D. -U.
Hwang, Phys. Rep. {\bf 424}, 175 (2006).

\bibitem{relation} J. Joo, and J. L. Leboitz, Phys. Rev. E {\bf 69}, 066105 (2004); R. Olinky, and L. Stone, Phys. Rev. E {\bf 70}, 030902 (2004); T. Zhou, J.-G. Liu, W.-J. Bai, G.-R. Chen, and B.-H. Wang,
Phys. Rev. E {\bf 74}, 056109 (2006); R. Yang, B.-H. Wang, J. Ren,
W.-J. Bai, Z.-W. Shi, W.-X. Wang, and T. Zhou, Phys. Lett. A {\bf
364}, 189 (2007).

\bibitem{p1} M. A. Jovanovic, \emph{Modeling large-scale
peer-to-peer networks and a case study of Gnutella} [M.S. Thesis],
University of Cincinnati (2001).

\bibitem{p2} \href{Http://www.gnutella.com}{Http://www.gnutella.com.}

\bibitem{SCN} F. Liljeros, C. R. Edling, L. A. N. Amaral, H. E. Stanley, and Y. \AA berg, Nature \textbf{411}, 907
(2001); W. -J. Bai, T. Zhou, and B. -H. Wang, Int. J. Mod. Phys. C
(to be published).

\bibitem{market} B. J. Kim, T. Jun, J. Y. Kim, and M. Y. Choi, Physica A \textbf{360},
493 (2005).

\bibitem{google} \href{Http://mail.google.com/mail/help/intl/en/about.html}{Http://mail.google.com/mail/help/intl/en/about.html.}

\bibitem{reason} Note that, if the connections of the underlying
networks are localized (e.g. lattices), then the mean-field
approximation is incorrect and the threshold value is not equal to
$1/A$.

\bibitem{EPJB1} M. Molloy, and B. Reed, Random Struct. Algorithms \textbf{6},
161 (1996); M. Molloy, and B. Reed, Combinatorics, Probab. Comput.
\textbf{7}, 295 (1998).

\bibitem{finite-size} R. Pastor-Satorras and A. Vespignani, Phys. Rev. E \textbf{65},
035108 (2002); D. -U. Hwang, S. Boccaletti, Y. Moreno, and R.
Lopez-Ruiz, Math. Biosci. \& Eng. {\bf 20}, 317 (2005).

\bibitem{21} B. Derrida and H. Flyvbjerg, J. Phys. A
\textbf{20} 5273 (1987).

\bibitem{random} J. M\"{u}ller, SIAM J. Appl. Math. {\bf 59}, 222 (1998); D. S. Callway, M. E. J. Newman, S. H. Strogatz,
and D. J. Watts, Phys. Rev. Lett. \textbf{85}, 5468 (2000); R.
Cohen, K. Erez, D. ben-Avraham, and S. Havlin, Phys. Rev. Lett.
\textbf{85}, 4626 (2000).

\bibitem{proportional} Z. Dezs\"{o}, and A.-L. Barab\'asi, Phys.
Rev. E {\bf 65}, 055103 (2002).

\bibitem{target} R. Pastor-Satorras and A. Vespignani, Phys. Rev.
E \textbf{65}, 036104 (2002).

\bibitem{target-net} Z. H. Liu, Y. C. Lai, and N. Ye, Phys. Rev. E
{\bf 67}, 031911 (2003); D. H. Zanette, and M. Kuperman, Physica A
{\bf 309}, 445 (2002); Y. C. Lai, Z. H. Liu, and N. Ye, Int. J.
Mod. Phys. B {\bf 17}, 4045 (2003); H. Zhang, Z. H. Liu, and W. C.
Ma, Chin. Phys. Lett. {\bf 23}, 1050 (2006).

\bibitem{target-model} N. Madar, T. Kalisky, R. Cohen, D.
ben-Avraham, and S. Havlin, Eur. Phys. J. B \textbf{38}, 269
(2004); T. Zhou, and B. -H. Wang, Chin. Phys. Lett. {\bf 22}, 1072
(2005); F. Takeuchi, and K. Yamamoto, Lect. Notes Comput. Sci.
{\bf 3514}, 956 (2005); W. -J. Bai, T. Zhou, and B. -H. Wang,
arXiv: physics/0610138.

\end{thebibliography}
\end{document}